\documentclass[10pt,journal,compsoc,a4paper]{IEEEtran}

\ifCLASSOPTIONcompsoc
  \usepackage[nocompress]{cite}
\else
  \usepackage{cite}
\fi

\usepackage{hyperref}

\ifCLASSINFOpdf
  \usepackage[pdftex]{graphicx}
\else
  \usepackage{graphicx}
\fi

\usepackage{url}

\usepackage{amsmath}
\usepackage{amssymb}

\begin{document}

\title{Arb: Efficient Arbitrary-Precision \\ Midpoint-Radius Interval Arithmetic}

\author{Fredrik~Johansson
\IEEEcompsocitemizethanks{\IEEEcompsocthanksitem F. Johansson was with
Inria Bordeaux-Sud-Ouest
and the University of Bordeaux, 33400 Talence, France.\protect\\
E-mail: fredrik.johansson@gmail.com}
\thanks{$ $}}


\IEEEtitleabstractindextext{%
\begin{abstract}
Arb is a C library for arbitrary-precision interval arithmetic
using the midpoint-radius representation, also known as ball arithmetic.
It supports real and complex numbers, polynomials, power series, matrices,
and evaluation of many special functions.
The core number types are designed for versatility and speed
in a range of scenarios, allowing performance that is competitive with
non-interval arbitrary-precision types such as MPFR and MPC floating-point numbers.
We discuss the low-level number representation, strategies for precision and error bounds,
and the implementation of efficient polynomial arithmetic with interval coefficients.
\end{abstract}

\begin{IEEEkeywords}
Arbitrary-precision arithmetic, interval arithmetic, floating-point arithmetic, polynomial arithmetic
\end{IEEEkeywords}}

\maketitle

\IEEEdisplaynontitleabstractindextext
\IEEEpeerreviewmaketitle

\IEEEraisesectionheading{\section{Introduction}\label{sec:introduction}}

\IEEEPARstart{I}{nterval} arithmetic allows computing
with real numbers in a mathematically rigorous
way by automatically tracking error bounds
through the steps of a program~\cite{tucker2011validated}.
Success stories of interval arithmetic in
mathematical research include Hales's proof of the
Kepler conjecture~\cite{hales2011revision},
Helfgott's proof of the ternary Goldbach conjecture~\cite{helfgott2015},
and Tucker's positive solution
of Smale's 14th problem concerning the existence
of the Lorenz attractor~\cite{tucker2002rigorous}.

The main drawback of interval arithmetic
is that the bounds can blow up catastrophically,
perhaps only telling us that $x \in [-\infty,\infty]$.
Assuming that all input intervals can be made
sufficiently precise, increasing the working precision is an effective way
to circumvent this problem.
One well-known implementation of arbitrary-precision
interval arithmetic is MPFI~\cite{RevolRouillier2005},
which builds on the MPFR~\cite{Fousse2007} library for
arbitrary-precision floating-point
arithmetic with correct rounding.
MPFI extends the principles of MPFR to provide a well-defined semantics
by guaranteeing that each built-in interval operation produces the
smallest possible output interval (of course, composing operations will
still generally lead to overestimation).
Due to the difficulty of computing optimal floating-point
enclosures,
MPFR, MPFI and the complex MPFR extension MPC~\cite{enge2011mpc} are
currently limited to a small set of built-in functions.

In this paper, we present Arb, a C library for arbitrary-precision
arithmetic using midpoint-radius intervals.
In midpoint-radius arithmetic, or ball arithmetic, a real number
is represented by an enclosure $[m \pm r]$ where the midpoint $m$
and the radius $r$ are floating-point numbers. The advantage
of this representation over the more traditional endpoint-based
intervals $[a,b]$ used in MPFI is that only $m$ needs to be tracked
to full precision; a few digits suffice for $r$, as in
$$\pi \in [3.14159265358979323846264338328 \pm 1.07 \cdot 10^{-30}].$$
At high precision, this costs $(1+\varepsilon)$ as much as
floating-point arithmetic, saving a factor two over endpoint intervals.

We argue that midpoint-radius arithmetic not only is a viable alternative to
endpoint-based interval arithmetic,
but competitive with floating-point arithmetic
in contexts
where arbitrary precision is used,
e.g.\ in computer algebra systems.
The small overhead of tracking errors automatically, if not completely negligible,
affords us the freedom to use more complex algorithms with
confidence in the output.

Our focus is on ``narrow'' intervals, say $[\pi \pm 2^{-30}]$;
that is, we are more
concerned with bounding arithmetic error starting from precise input
than bracketing function images on ``wide'' intervals, say $\sin([3,4])$.
For the latter job, high-degree Taylor approximations
are an alternative to direct application of interval arithmetic.
Arb has good support for Taylor expansion (automatic
differentiation), though presently only in one variable.

We use the ball representation for real numbers,
constructing complex numbers,
polynomials and matrices out of real balls.
This is the most convenient approach, but we note
that the concept of ball arithmetic can
be generalized directly to normed vector spaces,
e.g.\ giving disks for complex numbers and norm perturbation
bounds for matrices, which has some advantages~\cite{vdH:ball}.
Ball arithmetic in some form
is an old idea, previously used in e.g.\
Mathemagix~\cite{mathemagix} and iRRAM~\cite{Muller2001}.
Our contributions include low-level optimizations as well as
the scope of high-level features.

One of our goals is fast, reliable evaluation of
transcendental functions, which are needed
with high precision in many scientific applications~\cite{bailey2015high}.
Arb has rigorous implementations of elementary, complete
and incomplete gamma and beta, zeta, polylogarithm,
Bessel, Airy, exponential integral,
hypergeometric, modular, elliptic and other special functions
with full support for complex variables.
The speed is typically better than
previous arbitrary-precision software, despite
tracking error bounds.
The purpose of this paper is not to describe algorithms for particular
mathematical
functions (we refer to \cite{Johansson2015elementary,Johansson2014hurwitz,Johansson2016hypergeometric}). Instead, we focus on how the
core arithmetic in Arb facilitates implementations.

A preliminary report about Arb was presented in~\cite{Johansson2013arb}; however,
the core arithmetic has subsequently been rewritten
and many features have been added.
The present paper offers a more detailed view and
covers new developments.

\section{Features and example applications}

Arb is free software distributed under the
GNU Lesser General Public License (LGPL).
The public git repository is \url{https://github.com/fredrik-johansson/arb/}
and documentation is available at \url{http://arblib.org/}.
The code is thread-safe, written in portable C,
and builds in most common environments. An extensive test suite is included.

Arb depends on GMP~\cite{GMP} or the fork MPIR~\cite{MPIR}
for low-level bignum arithmetic, MPFR for some operations on floating-point
numbers and for testing (MPFR numbers are not used directly),
and FLINT \cite{Hart2010} for arithmetic over the exact rings
$\mathbb{Z}$, $\mathbb{Q}$ and $\mathbb{Z}/n\mathbb{Z}$ and polynomials
over these rings.
Conceptually, Arb extends FLINT's
numerical tower to the rings $\mathbb{R}$ and $\mathbb{C}$,
and follows similar coding conventions as FLINT.
Arb provides the following core types:
\begin{itemize}
\item \texttt{arf\_t} - arbitrary-precision floating-point numbers
\item \texttt{mag\_t} - unsigned floating-point numbers
\item \texttt{arb\_t} - real numbers, represented in midpoint-radius interval form $[m \pm r]$ where $m$ is an \texttt{arf\_t} and $r$ is a \texttt{mag\_t}
\item \texttt{acb\_t} - complex numbers, represented in Cartesian form $a+bi$ where $a,b$ are \texttt{arb\_t} real intervals
\item \texttt{arb\_poly\_t}, \texttt{acb\_poly\_t} - real and complex dense univariate polynomials
\item \texttt{arb\_mat\_t}, \texttt{acb\_mat\_t} - dense matrices
\end{itemize}

Each type comes with a set of methods.
For example, \texttt{arb\_add(z, x, y, prec)}
sets the \texttt{arb\_t} variable \texttt{z}
to the sum of the \texttt{arb\_t} variables
\texttt{x} and \texttt{y}, performing the computation at \texttt{prec} bits of precision.

In the git version as of November 2016,
there are around 1850 documented methods in total, including
alternative implementations of the same mathematical operation.
For example, there are methods
for computing the Riemann zeta function $\zeta(s)$
using Borwein's algorithm, the Euler product, Euler-Maclaurin summation,
and the Riemann-Siegel formula.
The user will most likely only need the ``top-level'' methods \texttt{arb\_zeta}, \texttt{acb\_zeta},
\texttt{arb\_poly\_zeta\_series} or \texttt{acb\_poly\_zeta\_series} (the
latter two compute series expansions, i.e.\ derivatives with respect to $s$)
which automatically try to choose the best algorithm depending on $s$ and
the precision, but methods for specific algorithms are available
for testing purposes and
as an option if the default choice is suboptimal.

Arb includes some 650 test programs that cover almost all the methods.
Typically, a test program exercises a single method (or variants of
the same method) by generating $10^3$ to $10^6$ random inputs,
computing the same mathematical quantity in two different ways (by using a functional
identity, switching the algorithm, or varying parameters such as the precision),
and verifying that the results are consistent, e.g.\ that two
intervals that should represent the same real number overlap. Random intervals
are generated non-uniformly to hit corner cases with high probability.

\subsection{Software and language issues}

C is a suitable language for library development
due to its speed, support for fine-grained memory management,
fast compilation, portability, and ease of interfacing
from other languages.
The last point is important, since
the lack of operator overloading and high-level generic
data types makes C cumbersome for many potential users.
High-level interfaces
to Arb are available in the Python-based SageMath computer algebra system~\cite{sage},
a separate Python module\footnote{\url{https://github.com/fredrik-johansson/python-flint}},
and the Julia computer algebra package Nemo\footnote{\url{http://nemocas.org}}.

Perhaps the biggest drawback of C as an implementation language
is that it provides poor protection
against simple programming errors. This makes stringent unit testing
particularly important.
We have found running unit tests with
Valgrind/Memcheck \cite{nethercote2007valgrind} to be indispensable for
detecting memory leaks, uses of uninitialized variables,
out-of-bounds array accesses, and other similar mistakes.

Arb is designed to be thread-safe, and in particular, avoids
global state.
However, thread-local storage is used for some internal caches.
To avoid leaking memory, the user should call \texttt{flint\_cleanup()}
before exiting a thread, which frees all caches used by FLINT,
MPFR and Arb.
A few Arb methods (such as matrix multiplication) can use several
threads internally, but only one thread is used by default; the user can set the number
of threads available for internal use with \texttt{flint\_set\_num\_threads()}.

\subsection{Numerical evaluation with guaranteed accuracy}

We now turn to demonstrating typical use.
With arbitrary-precision interval arithmetic,
a formula can often be evaluated to 
a desired tolerance by trying with few guard bits
and simply starting over with more guard bits if the resulting interval
is too wide.
The precision steps can be fine-tuned for a specific problem,
but generally speaking,
repeatedly doubling either the total precision or the guard bits
tends to give close to optimal performance.
The following program computes $\sin(\pi + e^{-10000})$
to a relative accuracy of 53 bits.

\begin{small}
\begin{verbatim}
#include "arb.h"
int main() {
    long prec;
    arb_t x, y;
    arb_init(x); arb_init(y);
    for (prec = 64; ; prec *= 2) {
        arb_const_pi(x, prec);
        arb_set_si(y, -10000);
        arb_exp(y, y, prec);
        arb_add(x, x, y, prec);
        arb_sin(y, x, prec);
        arb_printn(y, 15, 0); printf("\n");
        if (arb_rel_accuracy_bits(y) >= 53)
            break;
    }
    arb_clear(x); arb_clear(y);
    flint_cleanup();
}
\end{verbatim}
\end{small}

The output is:

\begin{small}
\begin{verbatim}
[+/- 6.01e-19]
[+/- 2.55e-38]
[+/- 8.01e-77]
[+/- 8.64e-154]
[+/- 5.37e-308]
[+/- 3.63e-616]
[+/- 1.07e-1232]
[+/- 9.27e-2466]
[-1.13548386531474e-4343 +/- 3.91e-4358]
\end{verbatim}
\end{small}

The Arb repository
includes example programs that use similar precision-increasing
loops to solve various standard test problems 
such as computing the
$n$-th iterate of the logistic map, the determinant of the $n \times n$
Hilbert matrix, or all the complex roots of a given degree-$n$ integer polynomial.

\subsubsection{Floating-point functions with guaranteed accuracy}

The example program shown above is easily turned into
a function that takes \texttt{double} input,
approximates some mathematical function to 53-bit accuracy,
and returns the interval midpoint rounded to a \texttt{double}.
Of course, the precision goal can be changed to any other
number of bits, and any other floating-point type can be used.

We have created a C header file that wraps Arb to
provide higher transcendental
functions for the C99 \texttt{double complex} type.\footnote{https://github.com/fredrik-johansson/arbcmath}
This code is obviously not competitive with optimized \texttt{double complex}
implementations, but few such implementations are available
that give accurate results on the whole complex domain.
The speed is highly competitive with other arbitrary-precision
libraries and computer algebra systems, many of which often
give wrong results. We refer to \cite{Johansson2016hypergeometric}
for benchmarks.

We mention a concrete use in computational hydrogeophysics:
Kuhlman\footnote{\url{https://github.com/klkuhlm/unconfined}} 
has developed a Fortran program for
unconfined aquifer test simulations,
where one model involves Bessel functions $J_{\nu}(z)$ and $K_{\nu}(z)$
with fractional~$\nu$ and complex~$z$.
Due to numerical instability in the simulation approach,
the Bessel functions are needed with
quad-precision (113-bit) accuracy.
A few lines of code are used to convert from Fortran quad-precision
types to Arb intervals, compute the Bessel functions accurately with Arb,
and convert back.

\subsubsection{Correct rounding}

We have developed an example program
containing Arb-based implementations of all the transcendental
functions available in version~3.1.3 of MPFR, guaranteeing
correct rounding to a variable number of bits in any of the MPFR supported
rounding modes (up, down, toward zero, away from zero, and to nearest
with ties-to-even) with correct detection of exact cases,
taking \texttt{mpfr\_t} input and output variables.
This requires approximately 500 lines of wrapper code in total for all functions.
The following simple termination test ensures that rounding the
midpoint of \texttt{x} to 53 bits in the round-to-nearest
mode will give the correct result
for this rounding mode:

\begin{verbatim}
if (arb_can_round_mpfr(x, 53, MPFR_RNDN))
    ...
\end{verbatim}

Correct rounding is more difficult than simply targeting a few ulps error,
due the table maker's dilemma.
Input where the function value is an exact floating-point number,
such as $x = 2^n$ for the function $\log_2(x) = \log(x) / \log(2)$,
would cause the precision-increasing loop to repeat forever
if the interval evaluation always produced
$[n \pm \varepsilon]$ with $\varepsilon > 0$.
Such exact cases are handled in the example program.
However, this code has not yet been optimized
for asymptotic cases where the function value is close to an exact floating-point number.
For example, $\tanh(10000) \approx 1$ to within 28852 bits.
MPFR internally detects such input and quickly returns either 1 or $1 - \varepsilon$
according to the rounding mode.
To compute $\tanh(2^{300})$, special handling is clearly necessary.
With the exception of such degenerate rounding cases,
the Arb-based functions generally run
faster than MPFR's built-in transcendental functions.
Note that the degenerate cases for correct rounding
do not affect normal use of Arb,
where correct rounding is not needed.

Testing the Arb-based implementations against their MPFR equivalents
for randomly generated inputs revealed cases where MPFR 3.1.3 gave
incorrect results for square roots, Bessel functions, and the
Riemann zeta function.
All cases involved normal precision and input values, which easily could
have occurred in real use.
The square root bug was caused by an edge case in bit-level
manipulation of the mantissa,
and the other two involved incorrect error analysis.
The MPFR developers were able to fix the bugs quickly,
and in response strengthened their test code.

The discovery of serious bugs in MPFR, a mature library
used by major applications such as SageMath and the GNU Compiler Collection (GCC),
highlights the need for peer review, cross-testing, and ideally,
computer-assisted formal verification of mathematical software.
Automating error analysis via interval arithmetic
can eliminate certain types of numerical bugs,
and should arguably be done more widely.
One must still have in mind that interval arithmetic is
not a cure for logical errors, faulty mathematical analysis, or bugs in the
implementation of the interval arithmetic itself.

\subsection{Exact computing}

In fields such as computational number theory and computational geometry,
it is common to rely on numerical
approximations to determine discrete information such as
signs of numbers.
Interval arithmetic is useful in this setting, since one can verify
that an output interval contains only points that are strictly positive
or negative,
encloses exactly one integer, etc.,
which then must be the correct result.
We illustrate with three examples from number theory.

\subsubsection{The partition function}

Some of the impetus to develop Arb
came from the problem of computing
the integer partition function
$p(n)$, which counts the number of ways~$n$ can be written as a sum
of positive integers, ignoring order.
The famous Hardy-Ramanujan-Rademacher formula
(featuring prominently in the plot of the
2015 film \textit{The Man Who Knew Infinity})
expresses $p(n)$ as an infinite
series of transcendental terms
\begin{equation}
p(n)= C(n) \sum_{k=1}^\infty \frac{A_k(n)}{k} \,
   I_{3/2} \left( \frac{\pi}{k} \sqrt{\frac{2}{3} \left(n - \frac{1}{24}\right)} \right),
\label{eq:hrr}
\end{equation}
where
$I_{3/2}(x) = (2/\pi)^{1/2} x^{-3/2} (x \cosh(x) - \sinh(x))$,
$C(n) = 2\pi (24n-1)^{-3/4}$,
and $A_k(n)$ denotes a certain complex exponential sum.
If a well-chosen truncation of~\eqref{eq:hrr} is evaluated using
sufficiently precise floating-point arithmetic, one obtains
a numerical approximation $y \approx p(n)$ such that $p(n) = \lfloor y + 1/2 \rfloor$.
Getting this right is far from trivial, as evidenced by the
fact that past versions of Maple computed
$p(11269), p(11566), \ldots$ incorrectly~\cite{oeismaple}.

It was shown in \cite{Johansson2012hrr} that
$p(n)$ can be computed in quasi-optimal time,
i.e.\ in time essentially linear in $\log(p(n))$,
by careful evaluation of \eqref{eq:hrr}.
This algorithm was implemented using MPFR arithmetic,
which required a laborious floating-point error analysis to
ensure correctness.
Later reimplementing the algorithm in Arb
made the error analysis nearly trivial and allowed improving speed by a factor
two (in part because of faster transcendental functions in Arb,
and in part because more aggressive optimizations could be made).

Arb computes the 111\,391-digit number $p(10^{10})$
in 0.3 seconds, whereas Mathematica 9.0 takes one minute.
Arb has been used
to compute the record value $p(10^{20}) = 1838176508 \ldots 6788091448$, an integer
with more than 11~billion
digits.\footnote{\url{http://fredrikj.net/blog/2014/03/new-partition-function-record/}}
This took 110 hours (205 hours split across two cores)
with 130~GB peak memory usage.

Evaluating \eqref{eq:hrr} is a nice benchmark
problem for arbitrary-precision software, because the logarithmic magnitudes
of the terms follow a hyperbola. For $n=10^{20}$,
one has to evaluate a few terms to billions of digits,
over a billion terms to low precision, and millions of terms to
precisions everywhere in between,
exercising the software at all scales.
For large $n$, Arb spends roughly half the time on computing $\pi$ and $\sinh(x)$ in
the first term of~\eqref{eq:hrr} to full precision.

The main use of computing $p(n)$ is to study residues $p(n)$ mod $m$,
so getting the last digit right is crucial.
Computing the full value of $p(n)$ via \eqref{eq:hrr} and then reducing
mod $m$ is the only known practical approach for huge $n$.

\subsubsection{Class polynomials}

The Hilbert class polynomial $H_D \in \mathbb{Z}[x]$ (where $D < 0$
is an imaginary quadratic discriminant) encodes
information about elliptic curves.
Applications of computing the coefficients of $H_D$ include
elliptic curve primality proving and generating curves with
desired cryptographic properties. An efficient way to construct $H_D$
uses the factorization
$$H_D = \prod_k (x - j(\tau_k))$$
where $\tau_k$ are complex algebraic numbers
and $j(\tau)$ is a modular function
expressible in terms of Jacobi theta functions.
Computing the roots numerically via the $j$-function
and expanding the product yields approximations of the
coefficients of $H_D$, from
which the exact integers can be deduced if
sufficiently high precision is used.
Since $H_D$ has degree $O(\sqrt{|D|})$ and
coefficients of size $2^{O(\sqrt{|D|})}$,
both the numerical evaluation of $j(\tau)$ and the polynomial arithmetic
needs to be efficient and precise for large $|D|$.
An implementation of this algorithm in Arb is as fast as the state-of-the-art
floating-point implementation by Enge~\cite{enge2009complexity},
and checking that each coefficient's computed interval
contains a unique integer gives a provably correct result.

\subsubsection{Cancellation and the Riemann hypothesis}

In \cite{Johansson2014hurwitz}, Arb was used to
rigorously determine values
of the first $n = 10^5$ Keiper-Li coefficients
and Stieltjes constants, which are certain
sequences of real numbers defined in terms of
high-order derivatives of the Riemann zeta function.
The Riemann hypothesis is equivalent to the statement
that all Keiper-Li coefficients $\lambda_n$ are positive,
and finding an explicit $\lambda_n < 0$
would constitute a disproof.
Unfortunately for the author, the data agreed
with the Riemann hypothesis
and other open conjectures.

These computations suffer from severe cancellation in the evaluated formulas,
meaning that to compute an $n$-th derivative to
just a few significant digits, or indeed just to determine its sign,
a precision of~$n$ bits has to be used;
in other words, for $n = 10^5$, Arb was
used to manipulate polynomials with $10^{10}$ bits of data.
Acceptable performance was possible thanks to
Arb's use of asymptotically fast polynomial arithmetic,
together with multithreading for parts of the computation that had
to use slower algorithms.

More recently, Arb has been used to study generalizations
of the Keiper-Li coefficients \cite{bucur2016li}.
Related to this example, Matiyasevich and Beliakov
have also performed investigations of Dirichlet L-functions
that involved using Arb to locate zeros to very high precision
\cite{beliakov2014}, \cite{beliakov2015approximation}.

\section{Low-level number types}

In Arb version 1.0, described in \cite{Johansson2013arb},
the same floating-point type was used for both
the midpoint and radius of an interval.
Since version 2.0, two different types are used.
An \texttt{arf\_t} holds an arbitrary-precision floating-point number (the midpoint),
and a \texttt{mag\_t} represents a fixed-precision error bound (the radius).
This specialization requires more code, but enabled
factor-two speedups
at low precision, with clear improvements
up to several hundred bits.
The organization of the data types is shown in Table~\ref{tab:datalayout}.
In this section, we explain the low-level design of the
\texttt{arf\_t} and \texttt{mag\_t} types
and how they influence \texttt{arb\_t} performance.

\subsection{Midpoints}

An \texttt{arf\_t} represents a dyadic number
$$a \cdot 2^b, \quad a \in \mathbb{Z}[\tfrac{1}{2}] \setminus \{0\}, \quad \tfrac{1}{2} \le |a| < 1, \quad b \in \mathbb{Z},$$
or one of the special values
$\{0, -\infty, +\infty, \operatorname{NaN}\}$.
Methods are provided for conversions, comparisons, and
arithmetic operations with correct directional rounding.
For example,
\begin{verbatim}
c = arf_add(z, x, y, 53, ARF_RND_NEAR);
\end{verbatim}
sets $z$ to the sum of $x$ and $y$, correctly rounded to the nearest
floating-point number with a 53-bit mantissa (with round-to-even on a tie).
The returned \texttt{int} flag $c$ is zero if the operation is exact,
and nonzero if rounding occurs.

An \texttt{arf\_t} variable just represents a floating-point value,
and the precision is considered a \emph{parameter of an operation}.
The stored mantissa~$a$ can have any bit length, and uses
dynamic allocation, much like GMP integers.
In contrast, MPFR stores the precision to be used for a result
as part of each \texttt{mpfr\_t} variable, and always allocates space
for full precision even if only a few bits are used.

The \texttt{arf\_t} approach is convenient for working with
exact dyadic numbers, in particular integers which can grow dynamically
from single-word values until they reach the precision limit
and need to be rounded. This is particularly useful for
evaluation of recurrence relations, in calculations
with polynomials and matrices, and in any situation where the inputs
are low-precision floating-point values but much higher precision
has to be used internally.
The working precision in an algorithm can also
be adjusted on the fly without changing each variable.

\begin{table}[t!]
\caption{Data layout of Arb floating-point and interval types.}
\label{tab:datalayout}
\begin{center}
\begin{tabular}{|l|l|r|}
\hline
\multicolumn{2}{|l|}{Exponent (\texttt{fmpz\_t})} & 1 word\phantom{s} \\
\hline
\multicolumn{2}{|l|}{Limb count + sign bit} & 1 word\phantom{s} \\
\hline
Limb 0 & Allocation count & 1 word\phantom{s} \\
\hline
Limb 1 & Pointer to $\ge\!3$ limbs & 1 word\phantom{s} \\
\hline
\multicolumn{2}{|l|}{\textbf{\texttt{arf\_t}}} & = 4 words \\
\hline
\end{tabular}
\end{center}

\begin{center}
\begin{tabular}{|l|l|r|}
\hline
\multicolumn{2}{|l|}{Exponent (\texttt{fmpz\_t})} & 1 word\phantom{s} \\
\hline
\multicolumn{2}{|l|}{Limb} & 1 word\phantom{s} \\
\hline
\multicolumn{2}{|l|}{\textbf{\texttt{mag\_t}}} & = 2 words \\
\hline
\end{tabular}
\end{center}

\begin{center}
\begin{tabular}{|l|l|r|}
\hline
\multicolumn{2}{|l|}{Midpoint (\texttt{arf\_t})} & 4 words \\
\hline
\multicolumn{2}{|l|}{Radius (\texttt{mag\_t})} & 2 words \\
\hline
\multicolumn{2}{|l|}{\textbf{\texttt{arb\_t}}} & = 6 words \\
\hline
\end{tabular}
\end{center}

\begin{center}
\begin{tabular}{|l|l|r|}
\hline
\multicolumn{2}{|l|}{Real part (\texttt{arb\_t})} & 6 words \\
\hline
\multicolumn{2}{|l|}{Imaginary part (\texttt{arb\_t})} & 6 words \\
\hline
\multicolumn{2}{|l|}{\textbf{\texttt{acb\_t}}} & = 12 words \\
\hline
\end{tabular}
\end{center}
\end{table}

\subsubsection{Mantissas}

The mantissa $\tfrac{1}{2} \le |a| < 1$ is stored as an array of words (limbs)
in little endian order, allowing GMP's \texttt{mpn} methods
to be used for direct manipulation.
Like MPFR's \texttt{mpfr\_t}, the mantissa is always normalized
so that the top bit of the top word is set.
This normalization makes addition slower than the unnormalized
representation used by GMP's \texttt{mpf\_t}, but it is more
economical at low precision and allows slightly faster multiplication.
For error bound calculations, it is also extremely convenient
that the exponent gives upper and lower power-of-two estimates.

The second word in the \texttt{arf\_t} struct
encodes a sign bit and the number of words
$n$ in the mantissa, with $n = 0$ indicating a special value.
The third and fourth words encode the mantissa.
If $n \le 2$, the these words store the limbs directly.
If $n > 2$, the third word specifies the number $m \ge n$ of allocated
limbs, and the fourth word is a pointer to $m$ limbs,
with the lowest $n$ being in use.
The mantissa is always normalized so that its
least significant limb is nonzero, and new space is allocated
dynamically if $n > m$ limbs need to be used.
If the number of used limbs shrinks to $n \le 2$,
the heap-allocated space is automatically freed.

On a 64-bit machine, an \texttt{arf\_t} with at most a
128-bit mantissa (and a small exponent) is represented entirely by a 256-bit
struct without separate heap allocation, 
thereby improving memory locality
and speeding up creation and destruction of variables,
and many operations use fast inlined code specifically for the $n \le 2$ cases.
When working at $p \ge 129$-bit precision, this design still
speeds up common special values such as all integers $|x| < 2^{128}$
and \texttt{double} constants, including the important special value zero.

In contrast, an \texttt{mpfr\_t} consists of four words (256 bits), plus
$\lceil p / 64 \rceil$ more words for the mantissa at $p$-bit precision which always
need to be allocated.
The MPFR format has the advantage of being slightly faster for
generic full-precision floating-point values, especially at
precision just over 128 bits, due to requiring less
logic for dealing with different lengths of the mantissa.


\subsubsection{Exponents}

The first word in the \texttt{arf\_t} struct represents
an arbitrarily large exponent as a FLINT integer, \texttt{fmpz\_t}.
An \texttt{fmpz\_t} with absolute value at most $2^{62}-1$ ($2^{30}-1$
on a 32-bit system) is immediate,
and a larger value encodes a pointer to a heap-allocated GMP bignum.
This differs from most other floating-point implementations,
including MPFR, where an exponent is confined to the numerical
range of one word.

Since exponents almost always will be small in practice, the
only overhead of allowing bignum exponents with this representation
comes from an extra integer comparison (followed by a predictable branch)
every time an exponent is accessed.
In fact, we encode infinities and NaNs using special exponent values
in a way that allows us to combine testing for large exponents
with testing for infinities or NaNs, which often must be done anyway.
In performance-critical functions where an input is used several times,
such as in a ball multiplication $[a \pm r] [b \pm s] = [ab \pm (|as| + |br| + rs)]$,
we only inspect each exponent once, and
use optimized code for the entire calculation
when all inputs are small.
The fallback code does not need to be optimized and can deal
with all remaining cases in a straightforward way
by using FLINT \texttt{fmpz\_t} functions to manipulate
the exponent values.

Using arbitrary-size exponents has two advantages.
First, since underflow or overflow cannot occur,
it becomes easier to reason about floating-point operations.
For example, no rewriting is needed to
evaluate $\sqrt{x^2 + y^2}$ correctly.
It is arguably easier for the user to check the exponent range \textit{a posteriori}
if the applications demands that it be bounded (e.g.\ if the goal is to emulate a hardware type)
than to work around underflow or overflow when it is unwanted.
Anecdotally, edge cases related to the exponent range have been
a frequent source of (usually minor) bugs in MPFR.

Second, arbitrary-size exponents become very useful when
dealing with asymptotic cases of special functions
and combinatorial numbers, as became clear
while developing \cite{mpmath}.
Typical quotients of large exponentials or gamma functions can be evaluated
directly without the need to make case distinctions
or rewrite formulas in logarithmic form (which can introduce
extra branch cut complications). Such rewriting may still be required
for reasons of speed or numerical stability (i.e.\ giving tight intervals),
but in some cases simply becomes an optional optimization.

Exponents can potentially grow so large that they slow down
computations or use more memory than is available.
We avoid this problem by introducing precision-dependent
exponent limits in relevant interval (\texttt{arb\_t} and \texttt{acb\_t}) functions,
where the information loss on underflow or overflow
gets absorbed by the error bound, as we discuss later.

\subsubsection{Feature simplifications}

The \texttt{arf\_t} type deviates from the IEEE~754 standard and
MPFR in a few important respects.

There is no global or thread-local state for exception flags,
rounding modes, default precision, exponent bounds, or other settings.
Methods that might round the output return a flag indicating whether the result is exact.
Domain errors such as division by zero or taking the square root
of a negative number result in NaNs which propagate through
a computation to allow detection at any later point.
Since underflow and overflow cannot occur at the level of floating-point
arithmetic, they do not need to be handled.
Memory allocation failure is considered fatal, and presumably
raises the process abort signal (provided that the system's \texttt{malloc}
allows catching failed allocations).
We claim that statelessness is a feature of good library design.
This allows referential transparency, and
it is arguably easier for the user to implement their own state than to be sure
that a library's state is in the wanted configuration at all times
(particularly since the library's state could be mutated by calls to external code
that uses the same library).

The set of methods
for the \texttt{arf\_t} type is deliberately kept small.
The most complicated methods are \texttt{arf\_sum}, which adds
a vector of floating-point numbers without intermediate rounding or overflow
(this is necessary for correct implementation of
interval predicates), 
and \texttt{arf\_complex\_mul} which computes
$(e+fi) = (a+bi)(c+di)$ with correct rounding.
Mathematical operations beyond addition,
multiplication, division and square roots of real numbers are
only implemented for the \texttt{arb\_t} type,
where correct rounding becomes unnecessary and
interval operations can be used internally to simplify the algorithms.

The \texttt{arf\_t} type
does not distinguish between positive and negative zero.
Signed zero is probably less useful in ball arithmetic
than in raw floating-point arithmetic.
Signed zero allows distinguishing between directional limits
when evaluating functions at discontinuities or branch cuts,
but such distinctions can be made
at a higher level without complicating the semantics of
the underlying number type.

With these things said, support for omitted IEEE~754 or MPFR features
could easily be accommodated
by the \texttt{arf\_t} data structure together with
wrapper methods.

\subsection{Radii and magnitude bounds}

The \texttt{mag\_t} type represents an
unsigned floating-point number $a \cdot 2^b$, $\tfrac{1}{2} \le a < 1$,
or one of the special values $\{0, +\infty\}$.
The mantissa $a$ has a fixed precision of 30 bits in order to
allow fast fused multiply-add operations on either 32-bit or 64-bit CPUs.
The arbitrary-size exponent $b$ is represented the same
way as in the \texttt{arf\_t} type.
Methods for the \texttt{mag\_t} type are optimized for speed, and may
compute bounds that are a few ulps larger
than optimally rounded upper bounds.
Besides being faster than an \texttt{arf\_t},
the \texttt{mag\_t} type allows cleaner code by 
by making upward rounding automatic
and removing the need for many sign checks.

A \texttt{double} could have been used instead of an integer mantissa.
This might be faster if coded carefully, though the need to normalize
exponents probably takes away some of the advantage.
We do some longer error bound calculations by temporarily
converting to \texttt{double} values, scaled so that overflow
or underflow cannot occur.
When using \texttt{double} arithmetic, we always 
add or multiply the final result by a small perturbation which can be
proved to give a correct upper bound in IEEE 754 floating-point
arithmetic regardless of the CPU rounding mode
or double-rounding on systems that use extended precision,
such at \texttt{x86} processors with the historical \texttt{x87} floating-point unit.
For correctness, we assume that unsafe rewriting of floating-point expressions
(e.g.\ assuming associativity) is disabled in the compiler,
and and we assume that certain \texttt{double} operations such
as \texttt{ldexp} and \texttt{sqrt} are correctly rounded.
As a side note, Arb sometimes uses the \texttt{libm}
transcendental functions in
heuristics (typically, for tuning parameters),
but never directly for error bounds.

\section{Arithmetic benchmarks}

Table~\ref{tab:timings} compares the performance of
Arb intervals (\texttt{arb\_t}),
MPFR 3.1.5 floating-point numbers (\texttt{mpfr\_t}) and MPFI 1.5.1 intervals (\texttt{mpfi\_t}) for
basic operations on real numbers.
Table~\ref{tab:timingscomplex} further
compares Arb complex intervals (\texttt{acb\_t}) and
MPC 1.0.3 complex floating-point numbers (\texttt{mpc\_t}).
An Intel i5-4300U CPU was used.

\begin{table}[!ht]
\caption{Time to perform a basic operation on intervals
with MPFI and Arb, normalized
by the time to perform the same operation on floating-point numbers (i.e.\ just the midpoints) with MPFR.
As operands, we take intervals for $x = \sqrt{3}, y = \sqrt{5}$ computed to full precision.}
\label{tab:timings}
\begin{center}
\renewcommand{\arraystretch}{1.2}
\begin{tabular}{c|cc|cc|cc}
prec  & MPFI & Arb & MPFI & Arb & MPFI & Arb \\ \hline
      & \multicolumn{2}{c|}{add} & \multicolumn{2}{c|}{mul} & \multicolumn{2}{c}{fma} \\ \hline
64    &  2.58 &  1.08 & 2.06 & 1.03 & 1.42 & 0.56 \\
128   &  2.15 &  1.03 & 2.16 & 1.09 & 1.62 & 0.68 \\
256   &  2.20 &  1.48 & 2.14 & 1.23 & 1.65 & 0.70 \\
1024  &  2.22 &  1.39 & 2.05 & 0.99 & 1.49 & 0.76 \\
4096  &  2.10 &  1.70 & 2.02 & 1.05 & 1.63 & 0.95 \\
32768 &  2.11 &  1.65 & 2.02 & 1.02 & 1.78 & 1.00 \\ \hline
      & \multicolumn{2}{c|}{div} & \multicolumn{2}{c|}{sqrt} & \multicolumn{2}{c}{pow} \\ \hline
64    & 2.96 & 1.72 & 2.02 & 1.78 & 0.97 & 0.09 \\
128   & 2.81 & 1.79 & 2.01 & 1.50 & 1.21 & 0.11 \\
256   & 2.56 & 1.38 & 2.15 & 1.31 & 1.40 & 0.13 \\
1024  & 2.23 & 0.92 & 2.03 & 1.09 & 1.68 & 0.29 \\
4096  & 2.09 & 0.82 & 2.03 & 1.04 & 1.94 & 0.67 \\
32768 & 1.98 & 1.01 & 2.02 & 1.04 & 1.95 & 0.79 \\
\end{tabular}
\end{center}
\end{table}

MPFI lacks fused multiply-add (fma) and pow operations, so we timed fma using
a mul followed by an add, and pow via log, mul and exp.
Unlike MPFI's built-in functions, these naive versions do not give
optimal enclosures.

Multiplication in Arb is about as fast as in MPFR, and twice as fast as
in MPFI. Ball multiplication
$[a \pm r] [b \pm s] = [ab \pm (|as| + |br| + rs)]$ requires four
multiplications and two additions (plus one more addition
bounding the rounding error in the midpoint multiplication $ab$),
but all steps except $ab$ are done with cheap $\texttt{mag\_t}$ operations.

Addition alone in Arb is slower than MPFR at high precision since
\texttt{arf\_add} is not as well optimized. However, addition is
not usually a bottleneck at high precision.
The fused multiply-add operation in Arb is optimized to be about
as fast as a multiplication alone at low to medium precision.
This is important for matrix multiplication and
basecase polynomial multiplication.
In the tested version of MPFR, a fused multiply-add is somewhat slower than two separate
operations, which appears to be an oversight and low-hanging
fruit for improvement.

Division and square root in Arb have high overhead at low precision
compared to MPFR,
due to the relatively complicated steps to bound the propagated error.
However, since the precision in these steps can be relaxed,
computing the bounds using \texttt{mag\_t}
is still cheaper than the doubled work to evaluate
at the endpoints which MPFI performs.

\begin{table}[!ht]
\caption{Time to perform a basic operation on complex intervals with Arb,
normalized by the time to perform the same operation on complex
floating-point numbers with MPC.
As operands, we take $x = \sqrt{3}+\sqrt{5}i, y = \sqrt{7}+\sqrt{11}i$.}
\label{tab:timingscomplex}
\begin{center}
\renewcommand{\arraystretch}{1.2}
\begin{tabular}{c|c|c|c|c|c|c}
prec  &  add  &  mul & fma & div & sqrt & pow \\ \hline
64    &  1.13 &  0.24 & 0.41 & 0.35 & 0.66 & 0.11 \\
128   &  1.50 &  0.29 & 0.41 & 0.34 & 0.77 & 0.11 \\
256   &  1.71 &  0.32 & 0.47 & 0.63 & 0.81 & 0.13 \\
1024  &  1.67 &  0.48 & 0.58 & 0.70 & 0.84 & 0.21 \\
4096  &  1.51 &  0.93 & 0.98 & 0.89 & 0.91 & 0.44 \\
32768 &  1.18 &  0.99 & 1.00 & 1.02 & 0.99 & 0.82 \\
\end{tabular}
\end{center}
\end{table}

The large speedup for the transcendental pow operation up to
about 4600 bits is due to the fast algorithm for
elementary functions described in \cite{Johansson2015elementary}.
At higher precision, Arb remains around 20\% faster than MPFR and MPC
due to a more optimized implementation of the
binary splitting algorithm to compute exp and atan.
Arb currently depends on MPFR for computing log, sin and cos above 4600 bits,
re-implementation of these functions being a future possibility.

As one more test of basic arithmetic, we consider
the following function that computes $N!$ given $a=0, b=N$.

\begin{verbatim}
void fac(arb_t res, int a, int b, int prec)
{
    if (b - a == 1)
        arb_set_si(res, b);
    else {
        arb_t tmp1, tmp2;
        arb_init(tmp1); arb_init(tmp2);
        fac(tmp1, a, a + (b - a) / 2, prec);
        fac(tmp2, a + (b - a) / 2, b, prec);
        arb_mul(res, tmp1, tmp2);
        arb_clear(tmp2); arb_clear(tmp2);
    }
}
\end{verbatim}

Table \ref{tab:timingsfact} compares absolute timings for
this code and the equivalent code using MPFR and MPFI.

\begin{table}[!ht]
\caption{Time in seconds to compute recursive factorial product with $N = 10^5$.}
\label{tab:timingsfact}
\begin{center}
\renewcommand{\arraystretch}{1.2}
\begin{tabular}{c|l|l|l}
prec  &  MPFR  &  MPFI & Arb \\ \hline
64    &  0.0129 &  0.0271 & 0.00315 \\
128   &  0.0137 &  0.0285 & 0.00303 \\
256   &  0.0165 &  0.0345 & 0.00396 \\
1024  &  0.0417 &  0.0852 & 0.00441 \\
4096  &  0.0309 &  0.0617 & 0.00543 \\
32768 &  0.109 &   0.234  & 0.00883 \\
\end{tabular}
\end{center}
\end{table}

In this benchmark, we deliberately allocate two
temporary variables at each recursion step.
The temporary variables could be avoided with a minor rewrite
of the algorithm,
but they are typical of real-world code.
Since most intermediate results are small integers, we also
see the benefit of allocating mantissas dynamically to optimize for
short partial results.
Computing $N!$ recursively is a model problem for various
divide-and-conquer tasks such as binary splitting evaluation
of linearly recurrent sequences.
The MPFR and MPFI versions could be optimized by
manually varying the precision or switching to integers at a certain
recursion depth (in fact, Arb does this in the computation of exp and atan
mentioned earlier), but this becomes inconvenient in more complicated problems,
such as the evaluation of the generalized hypergeometric series
${}_pF_q(a_1,\ldots,a_p; b_1,\ldots,b_q; z)$ where the parameters
(which may be complex numbers and even truncated power series)
can have mixed lengths and sizes.

\section{Precision and bounds}

By definition, interval arithmetic must preserve set inclusions. That is,
if $f$ is a point-valued function, $F$ is a valid interval
extension of $f$ if for any set $X$
and any point $x \in X$, the inclusion $f(x) \in F(X)$ holds.
This leaves considerable freedom in choosing what set $F(X)$ to
compute.

For basic \texttt{arb\_t} arithmetic operations, we generally
evaluate the floating-point
operation on the midpoint at $p$-bit precision, bound the propagated
error, and add a tight bound for the result of rounding the midpoint.
For example, addition becomes
$[m_1 \pm r_1] + [m_2 \pm r_2] = [\operatorname{round}_p(m_1 + m_2) \pm (r_1 + r_2 + \varepsilon_{\operatorname{round}})]$
where the radius operations are done with upward rounding.
In this case, the error bounds are essentially the best possible,
up to order $2^{-30}$ perturbations in the \texttt{mag\_t} radius operations.

For more complicated operations, the smallest possible enclosure
can be very difficult to determine. The design
of interval functions $F$ in Arb has largely been dictated by evaluation speed
and convenience, following the philosophy that ``good enough'' error bounds
can serve until a concrete application is found that mandates optimization.

\subsection{Generic error bounds}

Since the user inputs the precision $p$ as a parameter, we can
think of $F_p$ as a sequence of functions, and formulate
some useful properties that should hold.
Clearly, if $x$ is a single point, then $F_p(x)$ should
converge to $f(x)$ when $p \to \infty$,
preferably with error $2^{O(1)-p}$. It is also nice to
ensure $F_p(\{x\}) = \{f(x)\}$ for all sufficiently
large $p$ if $f(x)$ is exactly representable.
If $f$ is continuous near the point $x$ and
the sequence of input sets $X_p$ converge to $x$ sufficiently rapidly,
then $F_p(X_p)$ should converge to $f(x)$ when $p \to \infty$.
In particular, if $f$ is Lipschitz continuous and $X_p$ has radius $2^{O(1)-p}$,
then $F_p(X)$ should preferably have radius $2^{O(1)-p}$.

Let $X = [m \pm r]$ and assume that $f$ is differentiable.
A reasonable compromise between speed and accuracy
is to evaluate $f(m)$ to $p$-bit accuracy
and use a first-order error propagation bound:
$$\sup_{|t| \le r} |f(m+t) - f(m)| \le C_1 |r|, \quad C_1 = \sup_{|t| \le |r|} |f'(m+t)|.$$
Of course, this presumes that a good bound for $|f'|$ is available.
A bound on $|f|$ can be included if $r$ is large.
For example, for $m,r \in \mathbb{R}$,
$\sup_{|t| \le |r|} |\sin(m+t) - \sin(t)| \le \min(2, |r|)$.

In practice, we implement most operations by composing simpler
interval operations; either because
derivative bounds would be difficult to compute accurately and quickly,
or because the function composition is numerically stable
and avoids inflating the error bounds too much.
Ideally, asymptotic ill-conditioning
is captured by an elementary prefactor such as $e^z$ or $\sin(z)$,
whose accurate evaluation is delegated to
the corresponding \texttt{arb\_t} or \texttt{acb\_t} method.
Some case distinctions may be required
for different parts of the domain.
For instance, Arb computes the complex tangent as
\begin{align*}
 \tan(z) &=
  \begin{cases}
   \displaystyle
   \frac{\sin(z)}{\cos(z)}                      & \text{if } |\operatorname{mid}(\operatorname{im}(z))| < 1 \\[1em]
   \displaystyle
   i - \frac{2i \exp(2iz)}{1 + \exp(2iz)}       & \text{if } \operatorname{mid}(\operatorname{im}(z)) \ge 1 \\[1em]
   \displaystyle
   -i + \frac{2i \exp(-2iz)}{1 + \exp(-2iz)}    & \text{if } \operatorname{mid}(\operatorname{im}(z)) \le 1 \\
  \end{cases}
\end{align*}
When $|\operatorname{im}(z)|$ is large, the first formula is a
quotient of two large exponentials. This causes error bounds to blow up
in interval arithmetic, and for sufficiently large $|\operatorname{im}(z)|$,
overflow occurs.
The alternative formulas only compute small exponentials and add them
to numbers of unit magnitude, which is numerically stable
and avoids overflow problems.

In general, transcendental functions are computed
from some combination of functional equations
and finite approximations (e.g.\ truncated Taylor and asymptotic series),
using most of the ``tricks from the book''.
There are usually three distinct steps. Evaluation parameters
(e.g.\ the series truncation order; working precision to
compensate for cancellation) are first chosen using fast heuristics.
The finite approximation formula is then evaluated using interval
arithmetic. Finally, a rigorous bound for the truncation error
is computed using interval or \texttt{mag\_t} operations.

\subsection{Large values and evaluation cutoffs}

If $x$ is a floating-point number of size $|x| \approx 2^n$, then
computing
$\sin(x)$ or $\exp(x)$ to $p$-bit accuracy
requires $n+p$ bits of internal precision for
argument reduction, i.e.\ subtracting a multiple of $\pi$ or $\log(2)$ from $x$
(the floating-point approximation of $\exp(x)$ will also have an $n$-bit exponent).
This is clearly futile if $x = 2^{2^{100}}$.
It is feasible if $x = 2^{2^{35}}$, but in practice
computing billions of digits of $\pi$
is likely to be a waste of time.
For example, when evaluating the formula
\begin{equation}
\label{eq:logexpsum}
\log(x) + \sin(x) \exp(-x)
\end{equation}
we only need a crude bound for the sine and exponential to
get an accurate result if $x \gg 0$.
To handle different ranges of $x$ and $p$,
the user could make case
distinctions, but automatic cutoffs are useful when calculations become more
complex.

As a general rule, Arb restricts internal evaluation parameters
so that a method does at most $O(\operatorname{poly}(p))$ work,
independent of the input value. This prevents too much
time from being spent on branches in an evaluation tree that
may turn out not to be needed for the end result. It allows
a simple precision-increasing loop to be used for ``black box'' numerical evaluation
that can be terminated at any convenient point if it
fails to converge rapidly enough.
In other words, the goal is not to try to solve the problem at any cost,
but to fail gracefully and allow the user to try an alternative approach.

The cutoffs should increase in proportion to the precision so that
not too much time is wasted at low precision
on subexpressions that may turn out not to be needed,
but so that the precise values still can be computed by setting
the precision high enough.

For real trigonometric functions and exponentials,
Arb effectively computes
\begin{align*}
 \sin(x) &=
  \begin{cases}
   [\sin(x) \pm \varepsilon]        & \text{if } n \le \max(65536, 4p) \\
   [\pm 1]        & \text{if } n > \max(65536, 4p),
  \end{cases}
\end{align*}
\begin{align*}
 e^x &=
  \begin{cases}
   [e^x \pm \varepsilon]                            & \text{if } n \le \max(128, 2p) \\
   [0,2^{-2^{\max(128, 2p)}}]       & \text{if } n > \max(128, 2p) \text{ and } x < 0 \\
   [\pm \infty]                         & \text{if } n > \max(128, 2p) \text{ and } x > 0. \\
  \end{cases}
\end{align*}

The automatic overflow and underflow for $\exp(x)$ is certainly necessary
with arbitrary-size exponents,
but arbitrarily bad slowdown for a function such as $\sin(x)$ is a concern even
with single-word exponents, e.g.\ with MPFR and MPFI.
Evaluation cutoffs are useful even if the user only intends
to work with modest numbers, one reason being
that extremely large values can result
when some initial rounding noise
gets amplified by a sequence of floating-point operations.
It is better to pass such input through quickly than to stall the computation. 
Exponential or trigonometric terms that become
irrelevant asymptotically also appear in connection with
special functions. For example, the
right-hand side in the digamma function reflection formula
$$\psi(1-z) = \psi(z) + \pi \cot(\pi z)$$
with $z \in \mathbb{C}$ has the same nature as \eqref{eq:logexpsum}.
In Pari/GP 2.5.5 and Mathematica 9.0, numerically evaluating
$\psi(-10 + 2^{100} i)$ results in an overflow
(Maple 18 succeeds, however).
Version 0.19 of mpmath manages by using arbitrary-precision exponents,
but is unable to evaluate $\psi(-10 + 2^{2^{100}} i)$.
With Arb, computing at 53-bit precision gives
\begin{align*}
\psi(-10 + 2^{100} i) &= [69.3147180559945 \pm 3.12 \cdot 10^{-14}] \\
                      &+ [1.57079632679490 \pm 3.40 \cdot 10^{-15}] i
\end{align*}
and
\begin{align*}
\psi(-10 + 2^{2^{100}} i) &= [8.78668439483320 \cdot 10^{29} \pm 4.35 \cdot 10^{14}] \\
                          &+ [1.57079632679490 \pm 3.40 \cdot 10^{-15}] i.
\end{align*}
This works automatically since a numerically stable formula
is used to compute $\cot(\pi z)$ (like the formula for $\tan(z)$),
and in that formula, the tiny exponential automatically
evaluates to a power-of-two bound with a clamped exponent.

\subsection{Branch cuts}

Arb works with principal branches, following
conventions most common in
computer algebra systems. In particular, the complex logarithm satisfies
$-\pi < \operatorname{im}(\log(z)) \le \pi$,
and the phase of a negative real number is $+\pi$.
A convenience of using rectangular complex intervals instead of
disks is that it allows representing line segments along branch cuts
without crossing the cuts.
When intervals do cross branch cuts, the image of the principal branch
includes the jump discontinuity.
For example,
$$\log(-100 + [\pm 1] i) = [4.6052 \pm 7.99 \cdot 10^{-5}] + [\pm 3.15] i.$$

It would be tempting to pick an arbitrary branch, e.g.\
that of the midpoint, to avoid the discontinuity.
However, this would break formulas where the same branch choice is assumed
in two subexpressions and rounding perturbations could place
the midpoints on different sides of the cut.

It is up to the user to rewrite formulas
to avoid branch cuts when preserving continuity is necessary.
For example, to compute both square roots of a complex number
(in undefined order), one can use
$(\sqrt{z}, -\sqrt{z})$ if $\operatorname{re}(\operatorname{mid}(z)) \ge 0$
and $(i \sqrt{-z}, -i \sqrt{-z})$
if $\operatorname{re}(\operatorname{mid}(z)) < 0$.
Arb has limited support for working with non-principal
branches of higher special functions: the Gauss hypergeometric function
${}_2F_1$ has a branch cut on $(1,\infty)$, which is used by default,
but a method is available for continuous
analytic continuation of ${}_2F_1$ along an arbitrary path,
which may cross the normal placement of the branch cut.

\subsection{Decimal conversion}

While computations are done in binary and binary
is recommended for serialization,
human-readable decimal output is important for user interaction.
The method \texttt{arb\_printn(x, d, flags)},
given an \texttt{arb\_t} $x = [m \pm r]$,
a decimal precision $d \ge 1$, and default flags 0,
prints a decimal interval of the form
$[m' \pm r']$ where:
\begin{itemize}
\item $m'$ and $r'$ are exact decimal floating-point numbers,
\item $m'$ has at most $d$-digit mantissa; $r'$ has three digits,
\item $m'$ is nearly a correctly rounded representation of $x$: it is allowed to differ from $x$ by at most one unit in the last place
(if $x$ is accurate to fewer than $d$ digits, $m'$ is truncated accordingly),
\item $x \subseteq [m' \pm r']$ (the output radius $r'$ takes into account both the original error $r$ and any error resulting from the binary-to-decimal conversion).
\end{itemize}

For example, $x = [884279719003555 \cdot 2^{-48} \pm 536870913 \cdot 2^{-80}]$
(a 53-bit accurate enclosure of $\pi$)
is printed as $[3.141592653589793 \pm 5.61 \cdot 10^{-16}]$ with $d = 30$
and as $[3.14 \pm 1.60 \cdot 10^{-3}]$ with $d = 3$.
The brackets and $\pm r'$ are omitted if $m' = x$. If
less than one digit of $x$ can be determined, $m'$ is omitted,
resulting in a magnitude-bound output such as $[\pm 1.23 \cdot 10^{-8}]$.
(The typesetting in conventional mathematical notation
is a liberty taken in this paper;
the verbatim output is an ASCII string
with C-style floating-point literals such as \texttt{[3.14 +/- 1.60e-3]}.)

A method is also provided for parsing back from a string.
In general, a binary-decimal-binary or decimal-binary-decimal
roundtrip enlarges the interval. However, conversions in either direction
preserve exact midpoints (such as $x = 0.125$ with $d \ge 3$) whenever possible.

The implementations are simple: interval
arithmetic is used to multiply or divide out exponents, and the
actual radix conversions are performed on big integers, with
linear passes over the decimal strings for rounding and formatting.

\section{Polynomials, power series and matrices}

\begin{figure*}[!htb]
\centering
\includegraphics[scale=0.63]{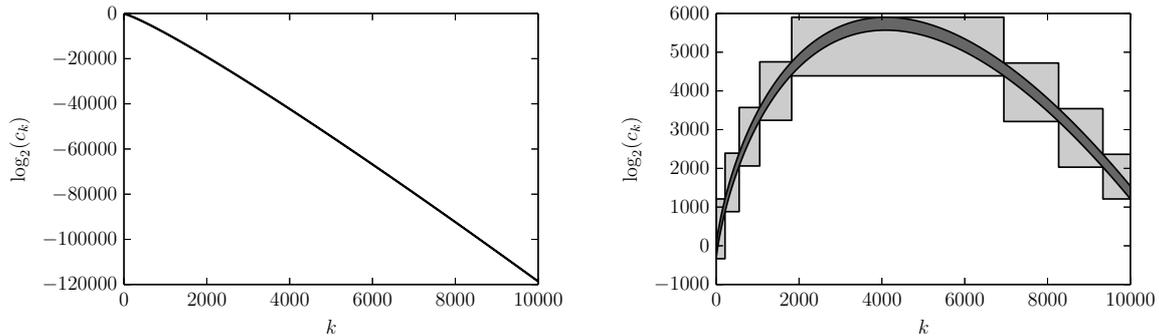}
\caption{Transformation used to square
$\exp(x) = \sum_k x^k / k! \in \mathbb{R}[x] / \langle x^n \rangle$
with $n = 10^4$ at $p = 333$ bits of precision.
The original polynomial, shown on the left, has an effective height of $\log_2(n!) + p \approx 119\,000$ bits.
Scaling $x \to 2^{12} x$ gives the polynomial on the right which
is split into 8 blocks of height at most 1511 bits, where the largest
block has a width of 5122 coefficients.}
\label{fig:digraph}
\end{figure*}

Arb provides matrices and univariate polynomials
with an eye toward computer algebra applications.
Polynomials are also used extensively within the library
for algorithms related to special functions.

Matrices come with rudimentary support for linear algebra,
including multiplication, powering,
LU factorization, nonsingular solving, inverse,
determinant, characteristic polynomial, and matrix exponential.
Most matrix operations currently
use the obvious, naive algorithms
(with the exception of matrix exponentials,
details about which are beyond the scope of this paper).
Support for finding eigenvalues is notably absent,
though computing roots of the characteristic polynomial
is feasible if the matrix is not too large.

Polynomials support all the usual operations
including arithmetic, differentiation, integration, evaluation, composition,
Taylor shift,
multipoint evaluation and interpolation,
complex root isolation, and reconstruction from given roots.
The polynomial types are also used to represent truncated power series,
and methods are provided for
truncated arithmetic, composition, reversion,
and standard algebraic and transcendental functions of power series.

Arb automatically switches between basecase algorithms for
low degree and asymptotically fast algorithms based on
polynomial multiplication for high degree.
For example, division, square roots and
elementary transcendental functions of power series use
$O(n^2)$ coefficient recurrences for short input
and methods based on Newton iteration
that cost $O(1)$ multiplications for long input.
Polynomial composition uses the divide-and-conquer algorithm~\cite{Hart2011practical},
and power series composition and reversion use
baby-step giant-step algorithms~\cite{BrentKung1978,Johansson2015reversion}.
Monomials and binomials are also handled specially in certain cases.

\subsection{Polynomial multiplication}

Since polynomial multiplication is the kernel
of many operations, it needs to be optimized for
both speed and accuracy, for input of all sizes and shapes.

When multiplying polynomials with interval coefficients, the $O(n^2)$
schoolbook algorithm essentially gives the best possible error bound
for each coefficient in the output (up to rounding errors
in the multiplication itself and under generic assumptions about
the coefficients).

The $O(n^{1.6})$ Karatsuba and $O(n \log n)$ FFT multiplication
algorithms work well when all input coefficients
and errors have the same absolute
magnitude, but they can give poor results when this
is not the case. The
effect is pronounced when manipulating power series with
decaying coefficients such as $\exp(x) = \sum_k x^k / k!$. Since the
FFT gives error bounds of roughly the same magnitude for all output
coefficients, high precision is necessary to produce accurate
high-order coefficients. Karatsuba multiplication
also effectively adds a term and then subtracts it again,
doubling the initial error, which leads to exponentially-growing
bounds for instance when computing
the powers $A, A^2, A^3, \ldots$ of a polynomial
via the recurrence $A^{k+1} = A^k \cdot A$.

We have implemented a version of the algorithm proposed by van der
Hoeven~\cite{vdH:stablemult} to combine numerical stability with FFT performance where
possible. This rests on several techniques:

\begin{enumerate}
\item Rather than directly multiplying polynomials with interval coefficients,
say $A \pm a$ and $B \pm b$ where $A, a, B, b \in \mathbb{Z}[\tfrac{1}{2}][x]$,
we compute $A B \pm (|A| b + a (|B| + b))$ using three multiplications
of polynomials with floating-point coefficients, where $|\cdot|$ denotes
the per-coefficient absolute value.
\item (Trimming: bits in the input coefficients that do not
contribute significantly can be discarded.)
\item Scaling: a substitution $x \to 2^c x$ is made to give
polynomials with more slowly changing coefficients.
\item Splitting: if the coefficients still vary too much, we write the
polynomials as block polynomials, say $A = A_0 + x^{r_1} A_1 + \ldots
x^{r_{K-1}} A_{K-1}$ and $B = B_0 + x^{s_1} B_1 + \ldots x^{s_{L-1}} B_{L-1}$, where
the coefficients in each block have similar magnitude. The block
polynomials are multiplied using $KL$ polynomial multiplications.
Ideally, we will have $K = L = 1$.
\item Exact multiplication: we finally use a fast algorithm to
multiply each pair of blocks $A_i B_j$. Instead of using
floating-point arithmetic, we compute $2^e A_i B_j \in \mathbb{Z}[x]$
exactly using integer arithmetic.
The product of the blocks is
added to the output interval polynomial using a single addition rounded to the
target precision.
\end{enumerate}

For degrees $n < 16$, we use the $O(n^2)$
schoolbook algorithm. At higher degree, we combine techniques 1 and
3-5 (technique 2 has not yet been implemented). We
perform a single scaling $x \to 2^c x$, where~$c$ is chosen
heuristically by looking at the exponents of the first and last
nonzero coefficient in both input polynomials and picking the weighted
average of the slopes
(the scaling trick is particularly effective when both $A$ and $B$ are power series
with the same finite radius of convergence).
We then split the inputs into blocks of height
(the difference between the highest and lowest exponent) at most $3 p
+ 512$ bits, where $p$ is the target precision.
The scaling and splitting is illustrated in Figure~\ref{fig:digraph}.

The exact
multiplications in $\mathbb{Z}[x]$ are done via FLINT. Depending on
the input size, FLINT in turn uses the schoolbook algorithm, Karatsuba,
Kronecker segmentation, or a Sch\"onhage-Strassen FFT.
The latter two algorithms have quasi-optimal bit complexity $\tilde{O}(n
p)$.

For the multiplications $|A| b$ and $a
(|B| + b)$ involving radii, blocks of width $n < 1000$ are processed using
schoolbook multiplication with hardware \texttt{double} arithmetic.
This has less overhead than working with
big integers, and guaranteeing
correct and accurate error bounds is easy since all coefficients
are nonnegative.

Our implementation follows the principle that polynomial
multiplication always should give error bounds of the same quality as
the schoolbook algorithm, sacrificing speed if necessary.
As a bonus, it preserves sparsity (e.g.\ even or odd polynomials)
and exactness of individual coefficients.

In practice, it is often the case that one needs $O(n)$ bits of precision
to compute with degree-$n$ polynomials and power series
regardless of the multiplication algorithm,
because the problems that lead to such polynomials
are inherently ill-conditioned.
In such cases, a single block will typically be used, so the
block algorithm is almost as fast as a ``lossy''
FFT algorithm that discards information about the smallest coefficients.
On the other hand, whenever low precision \emph{is} sufficient
with the block algorithm and a ``lossy'' FFT requires
much higher precision for equivalent output accuracy,
the ``lossy'' FFT is often even slower than the schoolbook algorithm.

Complex multiplication is reduced to four real multiplication
in the obvious way. Three multiplications would be sufficient
using the Karatsuba trick, but this suffers from the
instability problem mentioned earlier. Karatsuba multiplication
could, however, be used for the exact stage.

\subsection{Polynomial multiplication benchmark}

Enge's MPFRCX library~\cite{engempfrcx} implements
univariate
polynomials over MPFR and MPC coefficients
without control over the error.
Depending on size, MPFRCX performs
polynomial multiplication using
the schoolbook algorithm,
Karatsuba, Toom-Cook, or a numerical FFT.

Table~\ref{tab:polymultime1} compares MPFRCX and Arb
for multiplying real and complex polynomials
where all coefficients have roughly the same magnitude
(we use the real polynomials $f = \sum_{k=0}^{n-1} x^k / (k+1)$,
$g = \sum_{k=0}^{n-1} x^k / (k+2)$ and complex polynomials
with similar real and imaginary parts).
This means that MPFRCX's FFT multiplication
computes all coefficients accurately and that
Arb can use a single block.

The results show that multiplying via FLINT
generally performs significantly
better than a numerical FFT with high-precision coefficients.
MPFRCX is only faster for small~$n$ and very high precision,
where it uses Toom-Cook while Arb uses the schoolbook algorithm.

Complex coefficients are about four times slower than real
coefficients in Arb (since four real polynomial multiplications are used)
but only two times slower in MPFRCX (since a real FFT
takes half the work of a complex FFT).
A factor two could theoretically be saved in Arb's complex
multiplication algorithm
by recycling the integer transforms, but this would
be significantly harder to implement.

\begin{table}
\caption{Time in seconds to multiply polynomials of length $n$
with $p$-bit coefficients having roughly unit magnitude.}
\label{tab:polymultime1}
\begin{center}
\renewcommand{\arraystretch}{1.2}
\begin{tabular}{cc|cc|cc}
\multicolumn{2}{c}{ } & \multicolumn{2}{c}{Real} & \multicolumn{2}{c}{Complex} \\
$n$ & $p$ & MPFRCX & Arb & MPFRCX & Arb \\ \hline
10 & $100$ & 1.3e-5 & 6.9e-6 & 6.4e-5 & 3.5e-5 \\
10 & $1000$ & 3.1e-5 & 2.1e-5 & 1.8e-4 & 9.4e-5 \\
10 & $10^4$ & 3.6e-4 & 4.4e-4 & 0.0015 & 0.0021 \\
10 & $10^5$ & 0.0095 & 0.012 & 0.034 & 0.055 \\ \hline
100 & $100$ & 6.0e-4 & 1.3e-4 & 0.0020 & 5.6e-4 \\
100 & $1000$ & 0.0012 & 4.5e-4 & 0.0042 & 0.0019  \\
100 & $10^4$ & 0.012 & 0.0076 & 0.043 & 0.031  \\
100 & $10^5$ & 0.31 & 0.11 & 0.98 & 0.42 \\ \hline
$10^3$ & $100$ & 0.015 & 0.0022 & 0.025 & 0.0091 \\
$10^3$ & $1000$ & 0.029 & 0.0061 & 0.049 & 0.026 \\
$10^3$ & $10^4$ & 0.36 & 0.084 & 0.59 & 0.34 \\
$10^3$ & $10^5$ & 9.3 & 1.2 & 16 & 4.4 \\ \hline
$10^4$ & $100$ & 0.30 & 0.034 & 0.55 & 0.14 \\
$10^4$ & $1000$ & 0.63 & 0.19 & 1.1 & 0.82 \\
$10^4$ & $10^4$ & 8.0 & 1.2 & 14 & 4.6 \\
$10^4$ & $10^5$ & 204 & 13 & 349 & 50 \\ \hline
$10^5$ & $100$ & 2.9 & 0.54 & 5.4 & 2.0 \\
$10^5$ & $1000$ & 6.3 & 2.5 & 11 & 10  \\
$10^5$ & $10^4$ & 77 & 23 & 142 & 96 \\ \hline
$10^6$ & $100$ & 553 & 6.3 & 1621 & 23 \\
$10^6$ & $1000$ & 947 & 28 & 3311 & 103 \\
\end{tabular}
\end{center}
\end{table}

We show one more benchmark in Table~\ref{tab:polymultime2}.
Define
$$f_n = x(x-1)(x-2)\cdots(x-n+1) = \sum_{k=0}^n s(n,k) x^k.$$
Similar polynomials appear in series expansions
and in manipulation of differential and difference operators.
The coefficients $s(n,k)$ are the Stirling numbers of the first kind,
which fall of from size about $|s(n,1)| = (n-1)!$ to $s(n,n) = 1$.
Let $P = \max_k \log_2 |s(n,k)| + 64$.
Using a tree (binary splitting) to expand the product
provides an asymptotically fast way to generate $s(n,0), \ldots, s(n,n)$.
We compare expanding $f_n$ from the linear factors using:
\begin{itemize}
\item FLINT integer polynomials, with a tree.
\item MPFRCX, at 64-bit precision multiplying out one factor at a time,
and at $P$-bit precision with a tree.
\item Arb, one factor at a time at 64-bit precision, and then
at 64-bit precision and exactly (using $\ge P$-bit precision) with a tree.
\end{itemize}
Multiplying out iteratively one factor at a time is numerically stable, i.e.\
we get nearly 64-bit accuracy for all coefficients with both
MPFRCX and Arb at 64-bit precision.
Using a tree, we need $P$-bit precision
to get 64-bit accuracy for the smallest coefficients with MPFRCX, since
the error in the FFT multiplication depends on the largest term.
This turns out to be slower than exact computation with FLINT,
in part since the precision in MPFRCX does not automatically
track the size of the intermediate coefficients.

With Arb, using a tree gives nearly 64-bit accuracy for all coefficients
at 64-bit precision, thanks to the block multiplication algorithm.
The multiplication trades speed for accuracy, but when $n \gg 10^2$, the tree is still
much faster than expanding one factor at a time.
At the same time, Arb is about as fast as FLINT for exact computation
when $n$ is large, and can transition seamlessly between the
extremes.
For example, 4096-bit precision takes 1.8 s at $n = 10^4$ and 174 s at $n = 10^5$,
twice that of 64-bit precision.

\begin{table}
\caption{Time in seconds to expand falling factorial polynomial.}
\label{tab:polymultime2}
\begin{center}
\renewcommand{\arraystretch}{1.2}
\begin{tabular}{c|c|c|c|c|c|c}
$n$     &  FLINT & MPFRCX & MPFRCX & Arb & Arb & Arb \\
        &  exact,  & 64-bit,   & $P$-bit, & 64-bit  &  64-bit, & exact, \\
        &  tree  & iter.   & tree   &  iter. & tree & tree \\ \hline
$10$    & 4.8e-7 & 1.8e-5 & 1.8e-5 & 4.6e-6 & 2.7e-6 & 2.7e-6 \\
$10^2$  & 1.2e-4 & 1.1e-3 & 9.0e-4 & 4.8e-4 & 2.0e-4 & 2.4e-4 \\
$10^3$  & 0.030  & 0.10   & 0.35    & 0.049 & 0.0099   & 0.032 \\
$10^4$  & 5.9    & 10    & 386     & 4.8 & 0.85    & 5.9 \\
$10^5$  &         & 1540    &         & 515 & 85     &   \\
$10^6$  &         &         &         & & 8823     &      \\
\end{tabular}
\end{center}
\end{table}

\subsection{Power series and calculus}

Automatic differentiation together with fast polynomial arithmetic
allows computing derivatives
that would be hard to reach with numerical differentiation methods.
For example, if $f_1 = \exp(x)$, $f_2 = \exp(\exp(x))$, $f_3 = \Gamma(x)$,
$f_4 = \zeta(x)$, Arb computes $\{f_k^{(i)}(0.5)\}_{i=0}^{1000}$ to 1000 digits
in 0.0006, 0.2, 0.6 and 1.9 seconds respectively.

Series expansions of functions
can be used to carry out analytic operations such as
root-finding, optimization and integration with rigorous error bounds.
Arb includes code for isolating roots of real analytic functions
using bisection and Newton iteration.
To take an example from \cite{Johansson2016hypergeometric},
Arb isolates the 6710 roots of the Airy function $\operatorname{Ai}(x)$ on $[-1000,0]$
in 0.4~s and refines all roots to 1000 digits in 16~s.

Arb also includes code for integrating complex analytic
functions using the Taylor method,
which allows reaching 100 or 1000 digits with moderate effort. This code is
intended more as an example than for serious use.

\section{Conclusion}

We have demonstrated that midpoint-radius interval arithmetic
can be as performant as floating-point arithmetic in an arbitrary-precision
setting, combining asymptotic efficiency with low overhead.
It is also often easier to use.
The efficiency compared to non-interval software
is maintained or even improves when we move
from basic arithmetic to some higher operations
such as evaluation of special functions and polynomial manipulation,
since the core arithmetic enables using advanced algorithms for such tasks.

There is currently no accepted standard for how midpoint-radius
interval arithmetic should behave.
In Arb, we have taken a pragmatic approach which seems
to work very well in practice.
Arguably, fine-grained determinism (e.g.\ bitwise reproducible
rounding for individual arithmetic operations)
is much less important in interval
arithmetic than in floating-point arithmetic
since the quality of an interval result can be validated after
it has been computed.
This opens the door for many optimizations.
Implementing algorithms that give better error bounds efficiently
can itself be viewed as a performance optimization,
and should be one of the points for further study.

\ifCLASSOPTIONcompsoc
  \section*{Acknowledgments}
\else
  \section*{Acknowledgment}
\fi

The research was partially funded by
ERC Starting Grant ANTICS 278537 and Austrian Science Fund (FWF) grant Y464-N18.
Special thanks go to the people who have made contributions to Arb:
Bill Hart, Alex Griffing, Pascal Molin, and
many others who are credited in the documentation.

\bibliographystyle{IEEEtran}
\bibliography{references}

\end{document}